# Urban Twitter Networks and Communities: A Case Study of Microblogging in Athens

Tasos Spiliotopoulos

Madeira Interactive Technologies Institute
University of Madeira
Funchal, Portugal

Ian Oakley

School of Design and Human Engineering
UNIST
Ulsan, South Korea

**Abstract.** This paper examines the community formed by the Twitter users that used a city-level hashtag. In particular, we provide a network perspective of the city of Athens, Greece, as demonstrated by the analysis and visualization of the relevant Twitter hashtag data, in order to present both an overview and deeper insights at the microblogging practices of this geographic local network. Further analysis suggests that the Twitter community defined by the members of the network shows strong signs of a real-life community.

## I. Introduction

Social media offer ways for users to communicate, consume information and network with one another that complement, extend, and even substitute for the more traditional processes of offline personal communication. However, the fundamental nature of the communities formed in social media is unclear and the extent to which these collectives resemble their offline peers is an open research topic.

This paper examines the community formed by local Twitter hashtags at the city level. In particular, we provide a network perspective of the city of Athens, Greece, as demonstrated by the analysis and visualization of the relevant Twitter hashtag data. This visualization and analysis aims at providing both an overview and deeper insights at the microblogging practices of this geographic local network. The second contribution of this paper is the investigation of the extent to which the community formed resembles a real, physical community.

## II. Related Work

A large corpus of the relevant Twitter literature has focused on studying events. For example, Cheong and Cheong [1] performed a social network analysis of tweets during the 2010-11 Australian floods, in order to identify active players and their effectiveness in disseminating critical information. Yardi and boyd [2] showed a relationship between structural properties in the Twitter network and geographic properties in the physical world by tracking Twitter hashtags related to two local events. De Choudhury et al [3] examined Twitter activity around a range of events, and identified three core user categories; Organizations, Journalists/Media Bloggers (J/MB), and Ordinary Individuals (OI). Hu et al [4] examined a large corpus of neighborhood-specific tweets to find that 55% of them were about current events.

Other research efforts have focused in determining whether social media communities resemble real-life communities, and to what extent. Quercia et al. [5] examined a dataset of tweets from London-based Twitter accounts - focusing on topics, geography and emotions - to find signs of physical communities. Grudz et al. [6] examined a personal (i.e. egocentric) Twitter network to reveal that it exhibits characteristics typical of a community. The approach followed in the latter paper was based on the application of three notions of online communities. First, Anderson's [7] definition of an *imagined community* included three key elements: common language, temporality, and the absence of (organization around and under) high centers. Second, Jones [8] described a *virtual settlement* as a prerequisite for an online community, which needs to satisfy four conditions: interactivity, more than two communicators, a common public place where members can meet and interact, and sustained membership over time. It should be noted, however, that although a virtual settlement is a prerequisite for a virtual community, it does not guarantee its presence. Third, In order for the Twitter users to experience a *Sense of Community* (SoC), according to McMillan and Chavis [9] they need to feel that they belong to the community (membership), they can make a difference to the community (influence), they provide support and are supported by other members (integration and fulfillment of needs), and they share history, common places, time together and similar experiences (shared emotional connection).

## III. Data Collection

We employed the Twitter API to collect data about all the tweets across the duration of a month (from February 18th to March 19th 2011) that contained the name of the city of Athens, Greece as a hashtag. There are many cities around the world with the name of Athens apart from Athens, Greece, with the most

The work reported in this paper is supported by FCT research grant SFRH/BD/65908/2009.

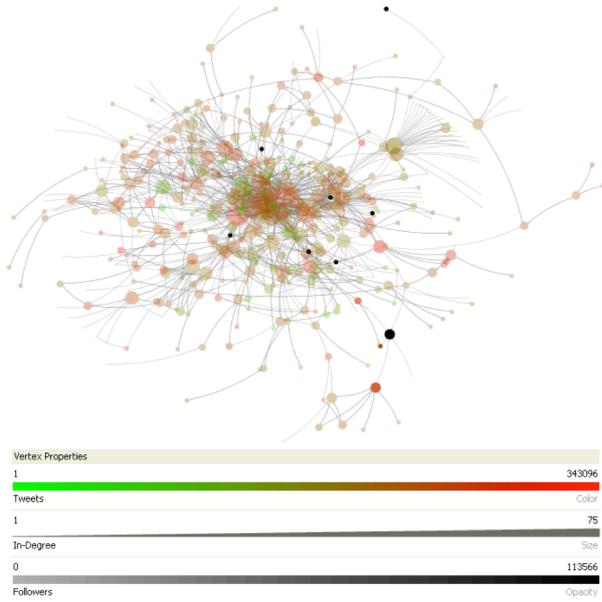

Fig. 1. The Athens F network.

prominent of them being Athens, Georgia, USA. The use of the hashtag "#athens -Georgia -AthensGA" as a search term ensured that tweets with the keywords "Georgia" and "AthensGA" were not included in our study. The rest of the instances that referred to other cities were cleaned manually. We gathered all the publicly accessible information for the account that made each tweet in the dataset and constructed three directed networks, based on the "follows" (F), "mentions and retweets" (M) and "replies to" (R) relationships among the accounts.

## IV. DESCRIPTION OF THE NETWORK

Overall, the F network consisted of 527 nodes (i.e. unique accounts that tweeted the hashtag) and 1947 edges (i.e. follow-type relationships among these accounts), the M network was much sparser with 416 nodes and 406 edges, and the R network was very small with only 17 nodes and 13 edges. In the remainder of the study we focus on the F network, since the M network is too sparse to have a meaningful structure and the R network is just too small.

Analysis of the network enabled the calculation of a number of metrics for each node, as well as some network-wide metrics. A node's *in-degree* is the number of incoming edges incident to the node. In this case, the number of followers (within our network) that an account has. A node's *out-degree* is the number of outgoing edges incident to the node. In this case, the number of accounts (from our network) that this account follows. *Graph density* is the ratio of the number of reported edges in the network divided by the number of all possible edges. The *number of connected components* refers to the distinct clusters within a network. *Geodesic distance* is the length of the shortest path between two nodes. The *average path length* is the average geodesic distance between all pairs of nodes in a network, while the *diameter* is the longest geodesic distance within the network, i.e. the maximum distance between two nodes. *Betweenness centrality* measures the extent to which other nodes lie in the geodesic path between pairs of nodes in the network. This metric generally concerns how other actors control or mediate the relations between dyads that are not directly connected, and as such it is an important indicator of control over information exchange or resource flows within a network. *Eigenvector centrality* is a measure of the importance of a node in a network and works by assigning relative scores to all nodes based on the principle that connections to high-scoring nodes contribute more to the score of the node in question than equal connections to low-scoring nodes. *PageRank* is a variation of the Eigenvector centrality algorithm. Finally, the *clustering coefficient* is a measure of the embeddedness of a node in its neighborhood. The average gives an overall indication of the clustering in the network, and high values are associated with a "small-world" effect.

TABLE I. METRICS OF THE ATHENS F NETWORK

| network size (no of nodes) | 527 | diameter | 11 |
|---|---|---|---|
| density | 0.007 | avg betweenness centrality | 1114 |
| connected components | 30 | avg eigenvector centrality | 0.002 |
| avg. geodesic distance | 4.17 | avg clustering coefficient | 0.211 |

Table 1 shows the values of the network-wide metrics in our network. These metrics provide a descriptive account of the network and also allow for the comparison of this network to other networks from the literature. Compared to similarly constructed networks about two geographically local events [2], the Athens network is 7-36 times denser and has 2-8 times higher in/out-degree. Of course, these values are even higher when comparing such networks with non-local hashtag networks [2]. Density implies a connected community of people who know a lot of other people in the community.

Figure 1 provides a visual representation of the network (the visualization was done with NodeXL [10]). This network has too many nodes to be able to identify the major players. It is evident, however, that there are nodes that are very central in the network. Additionally, there are nodes that have many followers and are very popular on Twitter in general (node opacity) that do not seem very important in our network. Nodes that are very active on Twitter in general (overall number of tweets – node color) tend to be more central in the network, a trend that is even stronger for nodes with high in-degree (node size).

TABLE II. TYPES OF ACCOUNT OF THE MOST CENTRAL NODES

|  | organizations | J/MB | OI | other |
|---|---|---|---|---|
| in-degree | 3 | 9 | 8 | 0 |
| out-degree | 2 | 7 | 11 | 0 |
| betweenness | 6 | 8 | 6 | 0 |
| pageRank | 0 | 9 | 11 | 0 |
| eigenvector | 4 | 8 | 7 | 1 |

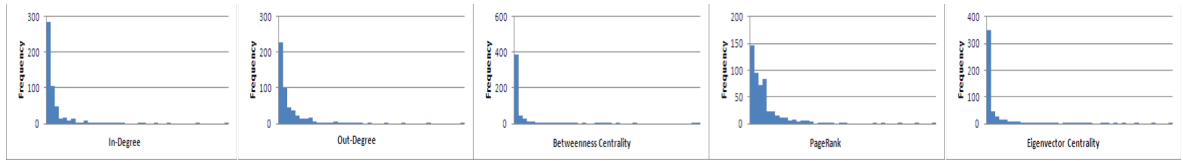

Fig. 2. Distribution of centrality measures.

In addition to the network-wide metrics, the distribution of values among the nodes in the network can have descriptive value. Figure 2 shows the distribution of the centrality measures in the network. As expected, with the exception of PageRank (which however is already defined in an exponential way), a few users are most connected or central and most are not well connected (or not connected at all) or peripheral in the network. We identified the 20 nodes with the highest value for each centrality measure and categorized them as one of the three core user categories suggested by the literature [3] (see Table 2). Compared to the composition of the user population tweeting about events [3], the journalists and media bloggers were highly overrepresented in our dataset, at the expense of ordinary individuals. It is not clear if this disparity is a result of the fact that we studied the most central nodes in the network, however.

## V. THE NETWORK AS A COMMUNITY

### A. Imagined Communities

According to Anderson's discussion [7] of imagined communities, one key element of community formation is **common language**. Indeed, as has been demonstrated by Gruzd et al [6], Twitter users make use of a common language. The fact that our sample is a minor subset of all twitter users does not affect this result. Another factor associated with the formation of imagined communities is **temporality**, i.e. the presence of a "homogeneous" sense of time. Twitter users are indeed expected to exhibit a continued imagined consciousness of a shared temporal dimension [6]. In our dataset, this is demonstrated by the large percentage of tweets that refer to current events, such as headline news, music concerts etc. Lastly, imagined communities are expected to refrain from being "organized around and under **high centers**" [7]. We consider betweenness centrality as the main network measure that describes the extent to which a node can be considered a "high center". Generally, removal of the nodes with high betweenness centrality from the network would lead to a big disruption in the network; both breaking up of the main connected component to smaller components, and increase in the geodesic distance between the nodes. In our dataset, we notice a power law distribution of the betweenness centralities. More particularly, the betweenness centrality of the node with the maximum value in the network is 27.9 times the average betweenness centrality of the network and 4338 times the median betweenness centrality. Thus, we conclude that there exist high centers in the networks under examination. However, as Grudz et al [6] point out, high centers play important roles in Twitter as community builders and information sources and this underlines an important difference between Twitter as an imagined community and the imagined communities in the physical world.

### B. Virtual Settlement

We consider the extent of mentioning and retweeting to be a measure of **interactivity** among the users, and thus we compare the M+R network to the F+M+R. This is a significantly different approach to measuring interactivity than the one followed in the study of personal Twitter networks [6], but is more applicable for complete networks like the one at hand. The edges of the M+R network account for 19.2% of the edges of the F+M+R network, while the vertices of the M+R network account for 75.6% of the vertices of the F+M+R network. This effectively means that 19.2% of the relations observed are conversational in nature (mentions and retweets), whereas 75.6% of all the users in the network engaged in conversational activities during the duration of the study. Given the relatively short duration of the study, we expected the first number to be quite low, since it compares the relations created during one month (through mentions and retweets) with those created during the whole lifetime of the accounts (through following). Also, we do not account for multiple retweets and mentions (we only count them once). Therefore, we consider 19.2% to be indicative of interactivity in the network, with the 75.6% value also strongly supporting this finding.

The network under study exhibits a **variety of communicators**, with hundreds of twitter users that used the hashtag. Moreover, there are users from many countries, accounts with different types of profiles and a variety of intended uses and gratifications. We believe that the network easily satisfies this requirement.

Also, by definition, the Athens network forms a **common public place where members can meet and interact**. Members of the network deliberately use a specific hashtag, so that other users looking for relevant information will follow them or search for this hashtag and retrieve this information.

It is very difficult to assess whether our network exhibits **sustained membership over time**, since we have collected data for a period of only one month. In order for a Twitter user to be included in the network, she would have to tweet one of the keywords. If the study lasted longer, then the network would only get larger since relatively few users actually delete their tweets. However, in order to capture the essence of sustained membership, we have to consider whether a Twitter user that is part of the network by being interested in a particular hashtag will continue to be interested after a certain time has passed. By studying the network we see that there are members that are directly associated with the city involved, such as local news agencies, local businesses, and local authorities. These accounts tend to be more central nodes in the network, while it is also probable that they will continue to post updates with the network hashtag. In addition, there are some personal Twitter accounts that seem to be interested in local information at a city level and we expect these accounts to continue being part of the community by posting updates, performing searches, or just following accounts that tweet these hashtags. Finally, there are users that are not expected to continue feeling part of the community after some time, such as tourists and visitors. These users will only tweet a certain hashtag (and thus become part of the network) once or twice when they visit the city, but not tweet again in the future.

*C. Sense of Community*

We already define the Twitter users that post updates with the hashtag of a specific city as **members** of the community of people that are interested in disseminating or receiving information about this particular city. In addition, we notice that, although there are many small connected components in the network, there exists a main connected component that contains a high percentage of the overall nodes and edges, something which shows that members are well connected, and strengthens the feeling of membership in the network. Particularly, 81.4% of the nodes and 91.5% of the edges belong to the main connected component of the network.

Another prerequisite for SoC is the ability of members of the community to **influence** one another. As measures of influence in our network we use retweeting and mentioning behavior, and betweenness centrality, since it has been demonstrated that the number of followers (i.e. in-degree) is not a good indicator of influence [11]. If a message is retweeted or a username is mentioned it means that some influence has been exerted to the user that retweets or mentions (i.e. they have been convinced of the importance of the tweet that they are retweeting). Thus, influence on Twitter is very tightly connected to interactivity, which we have already shown to be present in the network.

In order to assess the extent to which the Athens network provides **integration and fulfillment of needs**, we performed further analysis of the tweets to find that 42.9% of them contain one or more URLs and are, thus, explicitly informational in nature. This value is very high compared to a Twitter egocentric network [6] and suggests that people rely on Twitter not only to connect to other people, but also to connect to information.

According to McMillan and Chavis [9] having shared common places is one of the factors influencing **shared emotional connections** (the others being sharing history, spending time together and having similar experiences). Since the Athens network is topical in nature and content revolves around a specific hashtag which involves a common place, there is indication that the members of the network indeed share emotional connections to some extent.

VI. CONCLUSION AND FUTURE WORK

In contrast to the large body of literature that has focused on events, this study was designed with a broader scope and a less specific goal in mind. The rudimentary Twitter hashtag, which contained just the name of the city, aimed at eliciting tweets that encompass a broad range of the city's activity from an assortment of user types. Interestingly, the Twitter network of the city of Athens was found to be far denser than similarly constructed Twitter networks about specific local events. Furthermore, the Twitter community defined by the members of the network exhibited strong signs of a real-life community. This analysis expands Grudz and colleagues' [6] examination of a Twitter personal network as a community, by investigating the community based upon a shared interest on a real, offline community (i.e. the city).

Ongoing work includes the comparison of similarly collected data for two other cities in different countries in the same timeframe, in order to investigate possible differences in the communities formed, and comparison of the data for Athens with data collected using the same procedure 21 months later to illustrate the network's evolution over time.